# A New Full-diversity Criterion and Low-complexity STBCs with Partial Interference Cancellation Decoding

Lakshmi Prasad Natarajan and B. Sundar Rajan
Dept. of ECE, IISc, Bangalore 560012, India
Email: {nlp,bsrajan}@ece.iisc.ernet.in

*Abstract*—Recently, Guo and Xia gave sufficient conditions for an STBC to achieve full diversity when a PIC (Partial Interference Cancellation) or a PIC-SIC (PIC with Successive Interference Cancellation) decoder is used at the receiver. In this paper, we give alternative conditions for an STBC to achieve full diversity with PIC and PIC-SIC decoders, which are equivalent to Guo and Xia's conditions, but are much easier to check. Using these conditions, we construct a new class of full diversity PIC-SIC decodable codes, which contain the Toeplitz codes and a family of codes recently proposed by Zhang, Xu et. al. as proper subclasses. With the help of the new criteria, we also show that a class of PIC-SIC decodable codes recently proposed by Zhang, Shi et. al. can be decoded with much lower complexity than what is reported, without compromising on full diversity.

## I. INTRODUCTION

Space-Time Block Codes (STBCs) which can provide full diversity with low decoding complexity are important from an implementation point of view. Complex orthogonal designs (CODs) are known to provide real symbol-by-symbol ML decodability and thus have least ML decoding complexity [1], [2], [3]. These codes, however, suffer from low rates as the number of transmit antennas increases. As a remedy, quasi-orthogonal designs were proposed [4]. These codes achieve higher rate at the cost of higher ML decoding complexity. Single complex symbol or double real symbol ML decodable quasi-orthogonal STBCs were constructed in [5], [6] and [7]. In [8] and [9], the framework for multigroup ML decodable codes was given. An STBC is $g$-group ML decodable if the information symbols of the STBC can be partitioned into $g$ sets, such that each set of symbols can be ML decoded independent of other sets. As a result, the number of symbols that have to be jointly decoded is less and hence these codes have low complexity ML decoders. In [10], fast-decodable STBCs were introduced. These codes were not multigroup ML decodable, but they still have low ML decoding complexity. More fast-decodable codes were constructed in [11], [12].

All the codes discussed in the previous paragraph rely on ML decoders to achieve full diversity. As a result, their decoding complexities are still high, especially when the number of antennas or the rate is high. On the other hand, STBCs that give full diversity with linear receivers (Zero-Forcing (ZF) or Minimum Mean-Square-Error (MMSE) receivers) [13], [14], have lower decoding complexities, since each information symbol is decoded independently of other symbols, but suffer from low rates and performance. Recently, Guo and Xia [15], [16], introduced PIC and PIC-SIC decoders and gave sufficient conditions for an STBC to achieve full diversity under PIC and PIC-SIC decoding. The class of PIC decoders includes the ML decoder, ZF decoder and a number of other receivers with complexity and performance that lie in between those of ML and ZF.

Consider an STBC obtained from a *design* [17], $\mathbf{X} = \sum_{i=1}^{K} x_i A_i$ where, $x_i$ are the real information symbols, the linear dispersion matrices $A_i \in \mathbb{C}^{T \times N}$ are linearly independent over $\mathbb{R}$, $T$ is the delay and $N$ is the number of transmit antennas. The rate of such an STBC is $K/2T$ complex symbols per channel use (cspcu). A *grouping scheme* is a partition $\mathcal{I}_1, \ldots, \mathcal{I}_g$ of the set $\{1, \ldots, K\}$, where $\mathcal{I}_k$ are called *groups*. There is a corresponding partition of the information symbols into $g$ sets, where for $k = 1, \ldots, g$, the $k^{th}$ set of symbols is $\{x_j | j \in \mathcal{I}_k\}$. A PIC receiver decodes each set of symbols independently of other sets. In order to decode the $k^{th}$ group of symbols, a PIC decoder first implements a linear filter to eliminate the interference from symbols in all other groups and then decodes all the symbols of the $k^{th}$ group jointly. A PIC-SIC receiver uses successive interference cancellation along with PIC decoding. Let $n_{max} = max\{|\mathcal{I}_k| \mid k = 1 \ldots, g\}$. We say that the grouping scheme $\mathcal{I}_1, \ldots, \mathcal{I}_g$ leads to $n_{max}$-*real symbol PIC decoding* or $n_{max}$-*real symbol PIC-SIC decoding* when a PIC decoder or a PIC-SIC decoder is used respectively, since each step of the decoding process involves the joint decoding of at the most $n_{max}$ real symbols.

Using Coordinate Interleaving [6], full-diversity, rate $4/3$ double-real symbol (single complex symbol) PIC decodable STBCs were constructed in [18] for $2$ and $4$ antennas. A systematic design of STBCs leading to full diversity with PIC and PIC-SIC decoding was proposed in [19]. In [20], STBCs that have low PIC and PIC-SIC decoding complexity were constructed using Alamouti code [21] structure.

The contributions and organization of this paper are as follows.

- We propose alternative sufficient conditions for an STBC to achieve full diversity under PIC and PIC-SIC decoding. We show that these conditions are equivalent to the conditions given by Guo and Xia [16]. The criteria in [16] are difficult to check, whereas the new conditions can be checked easily. The use of the proposed criteria makes the problem of finding full-diversity codes easier (Section II).
- With the help of the new full-diversity conditions, for any number of antennas $N$ and any choice of $\lambda \leq N$, we construct full-diversity, $\lambda$-real symbol PIC-SIC decodable codes with rates arbitrarily close to $\lambda$ cspcu. This class of codes allows one to trade rate for decoding comfort. The proposed class of codes includes (see Table I and Fig. 1):
  - a family of codes from [19], but with a new choice of grouping scheme, leading to lower decoding complexities than those reported in [19],
  - the single real symbol PIC decodable Toeplitz codes [13],
  - the two antenna, rate $4/3$ code from [18].

  Specifically, for any choice of $\lambda \leq N \leq T$, we construct STBCs for $N$ antennas with delay $T$, rate $\lambda\left(1 - \frac{N-1}{T}\right)$ cspcu and worst-case PIC-SIC decoding complexity of $M^{\frac{\lambda-1}{2}}$, where, $M$ is the size of the complex constellation used. With large enough $T$, we get rates close to $\lambda$ (Section III).
- Using the new full-diversity criteria, we give a new grouping scheme for the full-diversity PIC-SIC decodable codes given in [20], with the number of real symbols per group only half of what is reported in [20]. The new grouping scheme, thus leads to huge reduction in decoding complexity. Specifically, this class is comprised of codes for any even values of $N$ and $T$ with $T \geq N$, having rate $\frac{N}{2}\left(1 - \frac{N-2}{T}\right)$ cspcu and worst-case PIC-SIC decoding complexity $M^{\frac{N-2}{4}}$. Whereas, the decoding complexity reported in [20] is $M^{\frac{N}{2}}$ (Section IV).

Directions for future work are discussed in Section V.

**Notation:** For a complex matrix $A$ the transpose, the conjugate and the conjugate-transpose are denoted by $A^T$, $\bar{A}$ and $A^H$ respectively. $||A||_F$ is the Frobenius norm of the matrix $A$. $I_n$ is the $n \times n$ identity matrix, $\mathbf{0}$ is the all zero matrix of appropriate dimension and $i = \sqrt{-1}$. The empty set is denoted by $\phi$. The cardinality of a set $\Gamma$ is denoted by $|\Gamma|$. The complement of a set $\Gamma$ with respect to a universal set $U$ is denoted by $\Gamma^c$, whenever $U$ is clear from context. For a square matrix $A$, $det(A)$ is the determinant of $A$. For a complex matrix $A$, $A_{Re}$ and $A_{Im}$ denote its real and imaginary parts respectively. Vectorization of a matrix $A$ is denoted by $vec(A)$ and the expectation operator is denoted by $\mathsf{E}(\cdot)$.

## II. A NEW FULL-DIVERSITY CRITERION

In this section, we give alternative conditions for an STBC to achieve full diversity with PIC and PIC-SIC decoding. These conditions are equivalent to the conditions given in [16], but are easier to check. This makes the problem of finding full-diversity PIC, PIC-SIC decodable codes and grouping schemes easier leading to low decoding complexity.

Consider an $N$ transmit antenna, $N_r$ receive antenna quasi-static Rayleigh flat fading MIMO channel given by $Y = \sqrt{\mathsf{SNR}}XH + W$, where $H$ is the $N \times N_r$ channel matrix, $X$ is the $T \times N$ matrix of transmitted signal, $W$ is the $T \times N_r$ additive noise matrix, $Y$ is the $T \times N_r$ matrix of received signal, all matrices being over the complex field $\mathbb{C}$, and SNR is the average signal-to-noise ratio at each receive antenna. It is assumed that $X$ takes values from a Space-Time Block Code (STBC) $\mathcal{C}$, satisfying the power constraint, $\mathsf{E}(||X||_F^2/T) = 1$. Let $\mathbf{X} = \sum_{i=1}^{K} x_i A_i$ be a design in $K$ real symbols $\{x_i\}$ with linear dispersion or weight matrices $A_i \in \mathbb{C}^{T \times N}$. The set of matrices $\{A_i\}$ must be linearly independent over $\mathbb{R}$. We obtain an STBC $\mathcal{C}(\mathbf{X}, \mathcal{A})$ from this design by letting the real symbols to take values from a signal set $\mathcal{A}$ which is a finite subset of $\mathbb{R}^K$, i.e., $\mathcal{C}(\mathbf{X}, \mathcal{A}) = \{\sum_{l=1}^{K} a_l A_l | [a_1, \ldots, a_K]^T \in \mathcal{A}\}$.

For a complex matrix $A$, define

$$\widetilde{vec}(A) = [vec(A_{Re})^T \ vec(A_{Im})^T]^T.$$

When using an STBC $\mathcal{C}(\mathbf{X}, \mathcal{A})$, the received signal $Y = \sqrt{\mathsf{SNR}}XH + W$ can be rewritten as

$$y = \widetilde{vec}(Y) = \sqrt{\mathsf{SNR}}Gx + \widetilde{vec}(W)$$

where, $G = G(H) = [\widetilde{vec}(A_1 H) \cdots \widetilde{vec}(A_K H)] \in \mathbb{R}^{2N_r T \times K}$ is a function of the channel realization $H$ and $x = [x_1, \ldots, x_K]^T \in \mathcal{A}$ is the vector of real information symbols. Let $\mathcal{I}_1, \ldots, \mathcal{I}_g$ be a grouping scheme such that, for each $k = 1, \ldots, g$, $|\mathcal{I}_k| = n_k > 0$ and $\mathcal{I}_k = \{i_{k,1}, \ldots, i_{k,n_k}\}$. Let $x_{\mathcal{I}_k} = [x_{i_{k,1}}, \ldots, x_{i_{k,n_k}}]^T$ denote the $k^{th}$ group of symbols. For $i = 1, \ldots, K$, let $g_i$ be the $i^{th}$ column of $G$. For $k = 1, \ldots, g$, define $G_{\mathcal{I}_k} = [g_{i_{k,1}} \cdots g_{i_{k,n_k}}]$ and $V_{\mathcal{I}_k} = span(\{g_j | j \notin \mathcal{I}_k\})$ is the subspace of $\mathbb{R}^{2N_r T}$ spanned by the set of vectors $\{g_j | j \notin \mathcal{I}_k\}$ over $\mathbb{R}$. Denote by $P_{\mathcal{I}_k}$ the matrix that projects a vector onto the subspace $V_{\mathcal{I}_k}^\perp$, the orthogonal complement of the subspace $V_{\mathcal{I}_k}$. Let $\tilde{V}_{\mathcal{I}_k} = span(\{g_j | j \in I_l, l > k\})$ and $\tilde{P}_{\mathcal{I}_k}$ be the matrix that projects a vector onto the subspace $\tilde{V}_{\mathcal{I}_k}^\perp$. It must be noted that $G$, $G_{\mathcal{I}_k}$, $V_{\mathcal{I}_k}$, $\tilde{V}_{\mathcal{I}_k}$, $P_{\mathcal{I}_k}$ and $\tilde{P}_{\mathcal{I}_k}$ are all functions of the channel realization $H$, although the notation we use does not explicitly show this aspect. However, we continue using this notation for the sake of brevity.

Assume that for each $k = 1, \ldots, g$, the vector symbols $x_{\mathcal{I}_k}$ are encoded independently of each other. If we define a permutation $\Pi$ of the coordinates of vectors in $\mathbb{R}^K$ as follows

$$\Pi(e_{i_{k,j}}) = e_{n_1 + \cdots + n_{k-1} + j} \text{ for all } k = 1, \ldots, g, \ 1 \leq j \leq n_k,$$

where $e_1, \ldots, e_K$ is the standard basis of $\mathbb{R}^K$, then $\Pi\mathcal{A} = \mathcal{A}_{\mathcal{I}_1} \times \cdots \times \mathcal{A}_{\mathcal{I}_g}$ where, $\mathcal{A}_{\mathcal{I}_k} \subset \mathbb{R}^{n_k}$.

A PIC decoder [15] with the grouping scheme $\mathcal{I}_1, \ldots, \mathcal{I}_g$ decodes each of the $g$ groups of symbols $x_{\mathcal{I}_k}$ as follows

$$\hat{x}_{\mathcal{I}_k} = arg \ min_{x_{\mathcal{I}_k} \in \mathcal{A}_{\mathcal{I}_k}} ||P_{\mathcal{I}_k} y - \sqrt{\mathsf{SNR}} P_{\mathcal{I}_k} G_{\mathcal{I}_k} x_{\mathcal{I}_k}||_F^2. \quad (1)$$

A PIC-SIC decoder [15] with the grouping scheme $\mathcal{I}_1, \ldots, \mathcal{I}_g$ decodes each of the $g$ groups of symbols $x_{\mathcal{I}_k}$ sequentially using the following algorithm. The decoder is initialized with $k = 1$ and $y_1 = y$.
- Step 1: Decode the $k^{th}$ vector of information symbols as

$$\hat{x}_{\mathcal{I}_k} := arg \ min_{x_{\mathcal{I}_k} \in \mathcal{A}_{\mathcal{I}_k}} ||\tilde{P}_{\mathcal{I}_k} y_k - \sqrt{\mathsf{SNR}} \tilde{P}_{\mathcal{I}_k} G_{\mathcal{I}_k} x_{\mathcal{I}_k}||_F^2. \quad (2)$$

- Step 2: Assign $y_{k+1} := y_k - \sqrt{\mathsf{SNR}} G_{\mathcal{I}_k} \hat{x}_{\mathcal{I}_k}$ and then assign $k := k+1$.
- Step 3: If $k > g$, stop. Else, go to Step 1.

Note that sphere-decoders [22] can be used to solve (1) and (2). The $k^{th}$ sphere-decoder jointly decodes $n_k$ real symbols or $\frac{n_k}{2}$ complex symbols. However, a sphere-decoder implementation of the ML decoder would jointly decode $K = \sum_{k=1}^{g} n_k$ real symbols. Thus, both PIC and PIC-SIC decoders have reduced average sphere-decoding complexities. The worst-case decoding complexity of both PIC and PIC-SIC decoders is $\sum_{k=1}^{g} M^{\frac{n_k}{2}}$, where $M$ is the cardinality of the underlying complex constellation. However, an STBC which does not have any additional property that can lead to low ML decoding complexity will have a worst-case ML decoding complexity of $M^{\sum_{k=1}^{g} \frac{n_k}{2}}$.

In [16], two sets of sufficient conditions were given for an STBC to achieve full-diversity, one each when the receiver employs a PIC and a PIC-SIC decoder respectively. The following theorem from [16], gives sufficient conditions for the STBC $\mathcal{C}(\mathbf{X}, \mathcal{A})$ to achieve full-diversity under PIC decoding. For any set of vectors $\mathcal{A}$, define $\Delta\mathcal{A} = \{a_1 - a_2 | a_1, a_2 \in \mathcal{A}\}$.

*Theorem 1 ([16]):* An STBC $\mathcal{C}(\mathbf{X}, \mathcal{A})$ achieves full-diversity under PIC decoding with a grouping scheme $\mathcal{I}_1, \ldots, \mathcal{I}_g$ if it satisfies the following two conditions:

1) $\mathcal{C}(\mathbf{X}, \mathcal{A})$ achieves full-diversity when an ML decoder is used and
2) for every $k = 1, \ldots, g$, every $H \neq \mathbf{0}$ and every $a_k \in \Delta\mathcal{A}_{\mathcal{I}_k} \setminus \{0\}$, we have $G_{\mathcal{I}_k} a_k \notin V_{\mathcal{I}_k}$.

We now provide an alternative condition for full-diversity under PIC decoding which is equivalent to the criterion of Theorem 1. Let $\Gamma = \{j_1, \ldots, j_{|\Gamma|}\}$ be any non-empty subset of $\{1, \ldots, K\}$ with $j_1 < j_2 < \cdots < j_{|\Gamma|}$. For any $u = [u_1, \ldots, u_{|\Gamma|}]^T \in \mathbb{R}^{|\Gamma|}$, define $X_\Gamma(u) = \sum_{i=1}^{|\Gamma|} u_i A_{j_i}$.

*Theorem 2:* An STBC $\mathcal{C}(\mathbf{X}, \mathcal{A})$ achieves full-diversity under PIC decoding with a grouping scheme $\mathcal{I}_1, \ldots, \mathcal{I}_g$, if it satisfies the following condition for every $k = 1, \ldots, g$:

- for every $a_k \in \Delta\mathcal{A}_{\mathcal{I}_k} \setminus \{0\}$ and every $u \in \mathbb{R}^{|\mathcal{I}_k^c|}$, we have: rank of $X_{\mathcal{I}_k}(a_k) + X_{\mathcal{I}_k^c}(u)$ is $N$.

Further, this condition is equivalent to the full-diversity criterion of Theorem 1.

*Proof:* It is enough to show that the criteria of Theorem 1 and Theorem 2 are equivalent. Let $k \in \{1, \ldots, g\}$, $\mathcal{I}_k = \{j_1, \ldots, j_{n_k}\}$ and $\mathcal{I}_k^c = \{l_1, \ldots, l_{K-n_k}\}$.

Let us assume that an STBC $\mathcal{C}(\mathbf{X}, \mathcal{A})$ satisfies the criteria posed in Theorem 1 under a grouping scheme $\mathcal{I}_1, \ldots, \mathcal{I}_g$. For any $H \neq \mathbf{0}$ and $a_k \in \Delta\mathcal{A}_{\mathcal{I}_k} \setminus \{0\}$ we have $G_{\mathcal{I}_k} a_k \notin V_{\mathcal{I}_k}$. Hence, for any $u = [u_1, \ldots, u_{N-n_k}]^T \in \mathbb{R}^{N-n_k}$ we have, $G_{\mathcal{I}_k} a_k + \sum_{i=1}^{K-n_k} u_i g_{l_i} \neq 0$. Since $g_q = \widetilde{vec}(A_q H)$ for $q = 1, \ldots, K$, we have,

$$\mathbf{0} \neq \sum_{i=1}^{n_k} a_i A_{j_i} H + \sum_{i=1}^{K-n_k} u_i A_{l_i} H = \left( X_{\mathcal{I}_k}(a_k) + X_{\mathcal{I}_k^c}(u) \right) H.$$

Since this is true for every $H \neq \mathbf{0}$, we have that no non-zero $N \times 1$ complex vector is orthogonal to all the columns of $(X_{\mathcal{I}_k}(a_k) + X_{\mathcal{I}_k^c}(u))^T$. Thus, the subspace spanned by the columns of $(X_{\mathcal{I}_k}(a_k) + X_{\mathcal{I}_k^c}(u))^T$ is the entire $\mathbb{C}^N$. Hence, the rank of $X_{\mathcal{I}_k}(a_k) + X_{\mathcal{I}_k^c}(u)$ is $N$ for every $u \in \mathbb{R}^{N-n_k}$.

Now assume that the rank of $X_{\mathcal{I}_k}(a_k) + X_{\mathcal{I}_k^c}(u)$ is $N$ for every $u \in \mathbb{R}^{N-n_k}$ and $a_k \in \Delta\mathcal{A}_{\mathcal{I}_k} \setminus \{0\}$. If $H \neq \mathbf{0}$, there must be at least one column of $H$ which is non-zero and hence has a non-zero dot product with at least one of the rows of $X_{\mathcal{I}_k}(a_k) + X_{\mathcal{I}_k^c}(u)$, since the rank of the row-space of $X_{\mathcal{I}_k}(a_k) + X_{\mathcal{I}_k^c}(u)$ is $N$, i.e., full. Thus,

$$\sum_{i=1}^{n_k} a_i A_{j_i} H + \sum_{i=1}^{K-n_k} u_i A_{l_i} H = \left( X_{\mathcal{I}_k}(a_k) + X_{\mathcal{I}_k^c}(u) \right) H \neq \mathbf{0}.$$

Thus, $G_{\mathcal{I}_k} a_k + \sum_{i=1}^{K-n_k} u_i g_{l_i} \neq 0$ for any $u \in \mathbb{R}^{K-n_k}$ and so $G_{\mathcal{I}_k} a_k \notin V_{\mathcal{I}_k}$. It only remains to show that such a code achieves full-diversity under ML decoding. Let $X_1$ and $X_2$ be two distinct codewords corresponding to distinct information symbol vectors $\xi_1, \xi_2 \in \mathcal{A}$ respectively. Since $a = \xi_1 - \xi_2 \neq 0$, there exists at least one $k \in \{1, \ldots, g\}$ such that, $a_{\mathcal{I}_k} \in \Delta\mathcal{A}_{\mathcal{I}_k} \setminus \{0\}$. Then, $X_1 - X_2 = X_{\mathcal{I}_k}(a_{\mathcal{I}_k}) + X_{\mathcal{I}_k^c}(a_{\mathcal{I}_k^c})$. Thus from the hypothesis, $X_1 - X_2$ has rank $N$. Thus the code achieves full diversity under ML decoding. ∎

The following theorem from [16] gives a sufficient condition for an STBC to achieve full diversity under PIC-SIC decoding.

*Theorem 3 ([16]):* An STBC $\mathcal{C}(\mathbf{X}, \mathcal{A})$ achieves full-diversity under PIC-SIC decoding with a grouping scheme $\mathcal{I}_1, \ldots, \mathcal{I}_g$ if it satisfies the following two conditions:

1) $\mathcal{C}(\mathbf{X}, \mathcal{A})$ achieves full-diversity when an ML decoder is used and
2) for every $k = 1, \ldots, g$, every $H \neq \mathbf{0}$ and every $a_k \in \Delta\mathcal{A}_{\mathcal{I}_k} \setminus \{0\}$, we have $G_{\mathcal{I}_k} a_k \notin \tilde{V}_{\mathcal{I}_k}$.

We now provide an alternative condition for full-diversity under PIC-SIC decoding, which is equivalent to the criterion of Theorem 3. For $k = 1, \ldots, g$, define $\tilde{\mathcal{I}}_k = \{j | j \in \mathcal{I}_l, l > k\}$.

*Theorem 4:* An STBC $\mathcal{C}(\mathbf{X}, \mathcal{A})$ achieves full-diversity under PIC-SIC decoding with a grouping scheme $\mathcal{I}_1, \ldots, \mathcal{I}_g$, if it satisfies the following condition for every $k = 1, \ldots, g$:

- for every $a_k \in \Delta\mathcal{A}_{\mathcal{I}_k} \setminus \{0\}$ and every $u \in \mathbb{R}^{|\tilde{\mathcal{I}}_k|}$, we have: rank of $X_{\mathcal{I}_k}(a_k) + X_{\tilde{\mathcal{I}}_k}(u)$ is $N$.

Further, this condition is equivalent to the full-diversity criterion of Theorem 3.

*Proof:* Proof is similar to the proof of Theorem 2. ∎

The new conditions, Theorems 2 and 4, are easier to check than the conditions of Theorems 1 and 3. This will be evident when we discuss codes and grouping schemes achieving full diversity under PIC and PIC-SIC decoding in Sections III and IV.

## III. A NEW CLASS OF FULL-DIVERSITY PIC-SIC DECODABLE CODES

In this section, for any integer $\lambda \geq 1$ and any number of antennas $N \geq \lambda$, we construct $\lambda$-real symbol PIC-SIC decodable codes with rates arbitrarily close to $\lambda$ cspcu. We then use the new criteria, Theorems 2 and 4, to show that these codes achieve full diversity with PIC-SIC decoding. The proposed class of codes includes a family of codes reported in [19]. However, we use a grouping scheme with double the number of groups reported in [19] and hence we show that these codes can be decoded with much lower complexities than those reported in [19]. The new class of codes also includes the rate 4/3 code for 2 antennas reported in [18].

*A. A New class of codes*

Consider integers $\lambda, n \geq 1$. Let the number of antennas $N \geq \lambda$, number of groups $g = 2n$ and number of real symbols $K = \lambda g = 2n\lambda$. For $k = 1, \ldots, g$, let the $k^{th}$ group be

$$\mathcal{I}_k = \{(k-1)\lambda + 1, (k-1)\lambda + 2, \ldots, k\lambda\}. \quad (3)$$

Each real symbol $x_i$, $i = 1, \ldots, K$, takes values from a regular PAM signal set, i.e., a finite subset of $\mathbb{Z}$, independent of other symbols. Clearly, the signal set $\mathcal{A} \subset \mathbb{R}^K$ is a cartesian product of $K$ one-dimensional real signal sets. Hence, the vectors of information symbols $x_{\mathcal{I}_1}, \ldots, x_{\mathcal{I}_g}$ are encoded independently of each other.

Let $Q \in \mathbb{R}^{\lambda \times \lambda}$ be a full-diversity rotation matrix [23] for the $\mathbb{Z}^\lambda$ lattice. For each $k = 1, \ldots, g$, define $z_{\mathcal{I}_k} = [z_{(k-1)\lambda+1}, z_{(k-1)\lambda+2}, \ldots, z_{k\lambda}]^T$ as $z_{\mathcal{I}_k} = Qx_{\mathcal{I}_k}$. For $m = 1, \ldots, n$, define $w_m \in \mathbb{C}^{\lambda \times 1}$ as follows:

$$w_m = [z_{(2m-2)\lambda+1} + iz_{(2m-1)\lambda+1}$$
$$z_{(2m-2)\lambda+2} + iz_{(2m-1)\lambda+2} \cdots z_{(2m-1)\lambda} + iz_{2m\lambda}]^T.$$

Note that $w_{m,Re}$ depends on symbols from $x_{\mathcal{I}_{2m-1}}$, and $w_{m,Im}$ depends on symbols from $x_{\mathcal{I}_{2m}}$. Since $N \geq \lambda$, there exist integers $d \geq 1$ and $r \in \{0, 1, \ldots, \lambda - 1\}$ such that $N = d\lambda + r$. For $m = 1, \ldots, n$, define vector $v_m \in \mathbb{C}^{N \times 1}$ as follows:

$$v_m = [w_m^T \ w_m^T \ \cdots \ w_m^T \ z_{(2m-2)\lambda+1} + iz_{(2m-1)\lambda+1}$$
$$\cdots z_{(2m-2)\lambda+r} + iz_{(2m-1)\lambda+r}]^T,$$

there being $d$ copies of $w_m^T$ in the above expression. Again, $v_{m,Re}$ depends on symbols from $x_{\mathcal{I}_{2m-1}}$, and $v_{m,Im}$ depends on symbols from $x_{\mathcal{I}_{2m}}$. Further, let $v_m = [v_m(1) \ v_m(2) \cdots v_m(N)]^T$ for complex scalars $v_m(1), \cdots, v_m(N)$. The proposed STBC is

$$\begin{bmatrix} v_1(1) & 0 & 0 & \cdots & 0 \\ v_2(1) & v_1(2) & 0 & \cdots & 0 \\ v_3(1) & v_2(2) & v_1(3) & \cdots & 0 \\ \vdots & \vdots & \vdots & \ddots & \vdots \\ \vdots & \vdots & \vdots & \cdots & v_1(N) \\ v_n(1) & v_{n-1}(2) & \cdots & \cdots & \vdots \\ 0 & v_n(2) & \cdots & \cdots & \vdots \\ \vdots & \vdots & \vdots & \cdots & \vdots \\ 0 & 0 & 0 & \cdots & v_n(N) \end{bmatrix}. \quad (4)$$

The delay of this code is $T = N + n - 1$. Consider the delay optimal case, i.e., $n = 1$. When $\lambda = N$, (4) reduces to a diagonal STBC which is 2 group ML decodable, the two groups being $x_{\mathcal{I}_1}$ and $x_{\mathcal{I}_2}$.

*B. Full-diversity*

Using the new criteria, Theorems 2 and 4, we show that the proposed STBCs achieve full diversity with PIC-SIC decoding in general, and PIC decoding in the case when $n = 2$.

*Proposition 1:* The family of STBCs (4) achieve full diversity with PIC-SIC group decoding and grouping scheme (3).

*Proof:* We use Theorem 4 to prove this proposition. Consider the case $k = 1$. The information symbols in $x_{\mathcal{I}_1}$ are encoded into the $N \times 1$ real vector $v_{1,Re}$. Since $Q$ is a full-diversity rotation for the $\mathbb{Z}^\lambda$ lattice, for any non-zero vector $a_{\mathcal{I}_1} \in \Delta\mathcal{A} \setminus \{0\}$, each coordinate of $v_{1,Re}$ is non-zero. Thus, for any choice of $v_{1,Im} \in \mathbb{R}^{N \times 1}$, each coordinate of $v_1$ is non-zero. Hence, for any choice of $v_{1,Im}, v_2, \ldots, v_n$, the resulting matrix has rank $N$. Hence, the matrix $X_{\mathcal{I}_1}(a_{\mathcal{I}_1}) + X_{\tilde{\mathcal{I}}_1}(u)$ has rank $N$ for any choice of $u \in \mathbb{R}^{K-\lambda}$. Thus, the condition of Theorem 4 is satisfied for $k = 1$. Using a similar argument for each $k = 2, \ldots, g$, it is straightforward to show that all the criteria of Theorem 4 are satisfied. Hence, the proposed code achieves full diversity with PIC-SIC decoding. ∎

*Proposition 2:* When $n = 1, 2$, the family of STBCs (4) achieve full diversity with PIC group decoding and grouping scheme (3).

*Proof:* Similar to the proof of Proposition 1, but uses Theorem 2 instead of Theorem 4. ∎

*C. Rate-Decoding Complexity-Delay tradeoff*

The class of codes proposed in this section have rate, $R = \frac{n\lambda}{N+n-1}$ cspcu for a given $n$, $\lambda$ and $N$. Equivalently, for any given $\lambda \geq 1$, $N \geq \lambda$ and $T \geq N$ we can choose $n = T - N + 1$ resulting in a $T \times N$ STBC with rate $R = \lambda\left(1 - \frac{N-1}{T}\right)$. By choosing $T$ large enough, a full-diversity, $\lambda$-real symbol PIC-SIC decodable code with rate $R$ arbitrarily close to $\lambda$ cspcu can be constructed using the given procedure. Thus, the single-real symbol PIC-SIC decodable codes of this section have rates arbitrarily close to 1 cspcu and the single-complex symbol (double real symbol) PIC-SIC decodable codes have rates arbitrarily close to 2 cspcu.

At each stage of PIC-SIC decoding (2) or PIC decoding (for the case $n = 2$) (1), $\lambda$ real symbols, $\{x_{(k-1)\lambda+1}, x_{(k-1)\lambda+2}, \ldots, x_{k\lambda}\}$ are jointly decoded. If $M$ is the cardinality of the underlying complex constellation, then each real symbol takes values from a $\sqrt{M}$-ary regular PAM signal set. For each of the $M^{\frac{\lambda-1}{2}}$ choices of values that the $\lambda - 1$ symbols $x_{(k-1)\lambda+2}, \ldots, x_{k\lambda}$ jointly assume, the value of $x_{(k-1)\lambda+1}$ that minimizes either (1) or (2) given the values of $x_{(k-1)\lambda+2}, \ldots, x_{k\lambda}$ can be found by simple scaling, rounding off and hard limiting. Thus, the order of worst case decoding complexity of the proposed codes is $M^{\frac{\lambda-1}{2}}$.

When, $N = 2$, $\lambda = 2$ and $T = 3$, we obtain the rate 4/3 code reported in [18], which has a worst-case PIC decoding complexity of $M^{0.5}$.

*Example 1:* Let $N = 3$, $\lambda = 2$ and $T = 6$. Corresponding value of $n$ is 4 and the code uses a PIC-SIC decoder with

$g = 8$ to obtain full diversity. The rate of the code is 4/3 cspcu and worst-case decoding complexity is $M^{0.5}$. The number of real symbols in the design is $K = 16$. Grouping scheme is: $\mathcal{I}_1 = \{1,2\}, \mathcal{I}_2 = \{3,4\}, \ldots, \mathcal{I}_8 = \{15,16\}$. The real symbols $z_j, j = 1, \ldots, 12$, are generated as:

$$[z_{2k-1}\ z_{2k}]^T = Q[x_{2k-1}\ x_{2k}]^T \text{ for } k = 1, \ldots, 8,$$

where, $Q$ is a $2 \times 2$ full-diversity rotation for $\mathbb{Z}^2$ lattice. The resulting STBC is

$$\begin{bmatrix} z_1 + iz_3 & 0 & 0 \\ z_5 + iz_7 & z_2 + iz_4 & 0 \\ z_9 + iz_{11} & z_6 + iz_8 & z_1 + iz_3 \\ z_{13} + iz_{15} & z_{10} + iz_{12} & z_5 + iz_7 \\ 0 & z_{14} + iz_{16} & z_9 + iz_{11} \\ 0 & 0 & z_{13} + iz_{15} \end{bmatrix}.$$

*Example 2:* Consider the case $N = \lambda = 4$ and $T = 6$. Corresponding value of $n$ is 3 and the code is decoded using a PIC-SIC decoder to get full diversity. The rate of this code is 2 cspcu and the worst-case decoding complexity is $M^{1.5}$. This stands in comparison with the rate 2, delay optimal, fast-ML-decodable code in [11], which has a worst-case ML decoding complexity of $M^{4.5}$ and rate 2 code in [12] with worst-case ML decoding complexity of $M^5$.

*Example 3:* Let $N = 4$, $\lambda = 3$ and $T = 9$. Corresponding value of $n$ is 6. Full diversity can be achieved using a PIC-SIC decoder. Rate of the code is 2 cspcu and the worst-case decoding complexity is $M$. Compared with the rate 2 code for 4 transmit antennas in Example 2, the code of this example has lower decoding complexity, but is of larger delay. This example illustrates the tradeoff between decoding complexity and delay that is achieved by the proposed class of codes.

### D. A family of codes in [19] as a subclass of proposed codes

A subclass of the proposed class of codes corresponding to the case $\lambda = N$ was first constructed in [19]. However, the worst-case decoding complexity of these codes was reported in [19] as $M^\lambda$ instead of the complexity $M^{\frac{\lambda-1}{2}}$ that we report in this paper. In [19], for each $m \in \{1, \ldots, n\}$, the symbols $x_{\mathcal{I}_{2m-1}}$ and $x_{\mathcal{I}_{2m}}$ constituted the $m^{th}$ group, even though they can be split into two groups without affecting the full-diversity property of the code.

### E. Toeplitz codes as a subclass of the proposed codes

Toeplitz codes [13] are known to provide full diversity with a zero-forcing receiver. The subclass of the codes proposed in this section corresponding to $\lambda = 1$ are exactly the Toeplitz codes with the underlying complex constellation being square QAM. In this case, the PIC decoder is nothing but a real symbol-by-symbol zero-forcing receiver. We now prove the full-diversity property using the new criterion in Theorem 2.

*Proposition 3:* For $\lambda = 1$ and any number of transmit antennas $N$, the STBC (4) with the grouping scheme (3) achieves full diversity with PIC decoding.

*Proof:* In this case, for every $m = 1, \ldots, n$, we have $v_m(1) = v_m(2) = \cdots = v_m(N) = x_{2m-1} + ix_{2m}$. Consider any $k \in \{1, \ldots, 2n\}$. Consider any real scalar $a_k \in \Delta \mathcal{A}_{\mathcal{I}_k} \setminus \{0\}$ and any $u \in \mathbb{R}^{2n-1}$. From Theorem 2, it is enough to show that the matrix $X_{\mathcal{I}_k}(a_k) + X_{\mathcal{I}_k^c}(u)$ is of rank $N$. Consider the smallest $l \in \{1, \ldots, n\}$ such that $v_l(1) \neq 0$. Such an $l$ always exists and $l \leq \lceil \frac{k}{2} \rceil$. This is so because, $v_{\lceil \frac{k}{2} \rceil, Re}(1)$ or $v_{\lceil \frac{k}{2} \rceil, Im}(1)$ is equal to $a_k$ when $k$ is odd or even respectively and hence $v_{\lceil \frac{k}{2} \rceil}$ is non-zero. Because of the choice of $l$, the first $l - 1$ diagonal layers will be zero and all entries in the $l^{th}$ diagonal layer will be non-zero. Thus, the $N \times N$ submatrix of $X_{\mathcal{I}_k}(a_k) + X_{\mathcal{I}_k^c}(u)$ consisting of all the $N$ columns and $N$ consecutive rows starting from the $l^{th}$ row will be lower triangular with non-zero, equal diagonal entries. Thus, this submatrix is of rank $N$ and hence the matrix $X_{\mathcal{I}_k}(a_k) + X_{\mathcal{I}_k^c}(u)$ is of rank $N$. ∎

## IV. A NEW GROUPING SCHEME FOR CODES IN [20]

In [20], systematic construction of STBCs which give full diversity with PIC and PIC-SIC decoding were given. These STBCs are constructed by replacing each element of an Alamouti code block [21] with a matrix containing multiple diagonal layers of coded symbols. With the help of the new full-diversity criteria, we propose a new grouping scheme for these codes with double the number of groups reported in [20]. Consequently, the new grouping scheme leads to huge reduction in decoding complexity. Finally, we compare the rate-decoding complexity pairs achievable by various PIC and PIC-SIC decodable codes available in the literature.

### A. A New grouping scheme

We now describe the codes proposed in [20] along with the new grouping scheme. Let the number of transmit antennas, $N$, be even. Let the number of real symbols per group be $\lambda = \frac{N}{2}$. Let $n \geq 1$ be any integer and the number of groups $g = 4n$. Number of real symbols in the design is $K = \lambda g = 2nN$. The new grouping scheme is as follows. For $k = 1, \ldots, g$, the $k^{th}$ group is

$$\mathcal{I}_k = \{(k-1)\lambda + 1, (k-1)\lambda + 2, \ldots, k\lambda\}. \quad (5)$$

Let each real symbol take values from a regular PAM constellation, i.e., a finite subset of $\mathbb{Z}$, independent of other symbols and let $Q \in \mathbb{R}^{\lambda \times \lambda}$ be a full-diversity rotation matrix for the integer lattice $\mathbb{Z}^\lambda$. For each $k = 1, \ldots, g$, define $z_{\mathcal{I}_k} = [z_{(k-1)\lambda+1}, z_{(k-1)\lambda+2}, \ldots, z_{k\lambda}]^T$ as $z_{\mathcal{I}_k} = Qx_{\mathcal{I}_k}$. For $m \in \{1, \ldots, n\}$ and $l \in \{1, \ldots, \lambda\}$, define $\mathfrak{A}(m,l)$ as in (6), given at the top of next page. $\mathfrak{A}(m,l)$ is an Alamouti code in real symbols $z_{(4m-4)\lambda+l}$, $z_{(4m-3)\lambda+l}$, $z_{(4m-2)\lambda+l}$ and $z_{(4m-1)\lambda+l}$. The STBC proposed in [20] upto a permutation

$$\mathfrak{A}(m,l) = \begin{bmatrix} z_{(4m-4)\lambda+l} + iz_{(4m-3)\lambda+l} & z_{(4m-2)\lambda+l} + iz_{(4m-1)\lambda+l} \\ -z_{(4m-2)\lambda+l} + iz_{(4m-1)\lambda+l} & z_{(4m-4)\lambda+l} - iz_{(4m-3)\lambda+l} \end{bmatrix}, \quad (6)$$

of rows and columns is

$$\begin{bmatrix} \mathfrak{A}(1,1) & \mathbf{0} & \cdots & \mathbf{0} \\ \mathfrak{A}(2,1) & \mathfrak{A}(1,2) & \cdots & \mathbf{0} \\ \vdots & \mathfrak{A}(2,2) & \ddots & \mathbf{0} \\ \vdots & \vdots & \ddots & \mathfrak{A}(1,\frac{N}{2}) \\ \vdots & \vdots & \cdots & \mathfrak{A}(2,\frac{N}{2}) \\ \vdots & \vdots & \cdots & \vdots \\ \mathfrak{A}(n,1) & \vdots & \cdots & \vdots \\ \mathbf{0} & \mathfrak{A}(n,2) & \cdots & \vdots \\ \vdots & \vdots & \ddots & \vdots \\ \mathbf{0} & \mathbf{0} & \cdots & \mathfrak{A}(n,\frac{N}{2}) \end{bmatrix}. \quad (7)$$

The STBC (7) consists of $n$ diagonal layers. Each diagonal layer has $\lambda = \frac{N}{2}$ Alamouti blocks that together encode $2N$ real symbols. The $2N$ real symbols can be divided into $4$ encoding groups each containing $\frac{N}{2}$ symbols. The four groups encoded by the $m^{th}$ layer are $x_{\mathcal{I}_{4m-3}}$, $x_{\mathcal{I}_{4m-2}}$, $x_{\mathcal{I}_{4m-1}}$ and $x_{\mathcal{I}_{4m}}$. The delay of the STBC (7) is $T = N + 2(n-1)$. For the delay optimal case, i.e., $n = 1$, (7) reduces to the 4-group ML decodable Precoded Coordinate Interleaved Orthogonal Design (PCIOD) given in [24].

In [20], full diversity was proved for a grouping scheme with $2n$ groups, which is only half the number of groups in new the grouping scheme. In terms of the new groups, the groups proposed in [20] are:

$$\mathcal{I}_1 \cup \mathcal{I}_2, \ \mathcal{I}_3 \cup \mathcal{I}_4, \ \cdots, \mathcal{I}_{4n-1} \cup \mathcal{I}_{4n}.$$

### B. Full-diversity

With the help of the new full-diversity criteria, Theorems 2 and 4, we now show that the STBC (7) yields full-diversity with the new grouping scheme.

*Proposition 4:* The family of STBCs (7) along with the grouping scheme (5) achieve full-diversity with PIC-SIC decoding.

*Proof:* Consider the case $k = 1$. The information symbol vector $x_{\mathcal{I}_1}$ is encoded into $z_{\mathcal{I}_1}$. The $\lambda$ coordinates of $z_{\mathcal{I}_1}$ act as one of the 4 real symbols in each of the $\lambda$ Alamouti blocks $\mathfrak{A}(1,1)$, $\mathfrak{A}(1,2)$,...,$\mathfrak{A}(1,\lambda)$ respectively. Since $Q$ is a full-diversity rotation for the integer lattice, for any $x_{\mathcal{I}_1} \in \Delta \mathcal{A}_{\mathcal{I}_k} \setminus \{0\}$, each of the $\lambda$ coordinates of $z_{\mathcal{I}_1}$ is non-zero. Hence, for any $z_{\mathcal{I}_k} \in \mathbb{R}^\lambda$, $k > 1$, each of the matrices $\mathfrak{A}(1,1)$, $\mathfrak{A}(1,2)$,...,$\mathfrak{A}(1,\lambda)$ is of full-rank. The determinant of the submatrix of $X_{\mathcal{I}_1}(z_{\mathcal{I}_1}) + X_{\tilde{\mathcal{I}}_1}(u)$ for any $u \in \mathbb{R}^{K-\lambda}$ consisting of the first $N$ rows and all the $N$ columns is the product $\prod_{l=1}^{\lambda} det(\mathfrak{A}(1,l)) \neq 0$. Hence, the matrix $X_{\mathcal{I}_1}(z_{\mathcal{I}_1}) + X_{\tilde{\mathcal{I}}_1}(u)$ is of rank $N$ for any $u \in \mathbb{R}^{K-\lambda}$. Using a similar argument for each $k = 1, \ldots, g$, we see that the STBC (7) satisfies the hypothesis of Theorem 4 for the grouping scheme (5) and hence achieves full diversity with PIC-SIC decoding. ∎

*Proposition 5:* When $n = 1, 2$, the family of STBCs (7) along with the grouping scheme (5) achieve full-diversity with PIC decoding.

*Proof:* Similar to the proof of Proposition 4, but uses Theorem 2 instead of Theorem 4. ∎

### C. Rate and Decoding complexity

For even number of transmit antennas, $N$, and code parameter $n \geq 1$, we have the number of real symbols $K = 2nN$. The rate of the code is $R = \frac{N}{2}\left(1 - \frac{N-2}{T}\right)$ cspcu, where both $N$ and $T$ are even and $T \geq N$. During PIC or PIC-SIC decoding, the number of real symbols that are jointly decoded is $\frac{N}{2}$. When a sphere-decoder is employed to solve (1) or (2), the dimension of the sphere-decoding problem is only $N/2$ over $\mathbb{R}$. This is in comparison with $N$-dimensional sphere-decoder employed in [20]. Thus, with the new grouping scheme, the average complexity of the decoder is reduced.

We now derive the worst-case decoding complexity when the new grouping scheme is employed. For each step of the decoding process (1) and (2), $N/2$ real symbols have to be jointly decoded. If $M$ is the cardinality of the underlying complex constellation, then each real symbol takes values from $\sqrt{M}$-ary regular PAM signal set. By jointly fixing the values of $N/2-1$ real symbols, the value of the last real symbol that minimizes (1) or (2) can be found out by scaling, rounding-off and hard limiting. The number of realizations of the set of $N/2-1$ real symbols is $M^{\frac{N/2-1}{2}}$. Thus, the order of the worst-case decoding complexity is $M^{\frac{N-2}{4}}$. This is much smaller than the complexity $M^{\frac{N}{2}}$ reported in [20].

The class of codes discussed in this section include the following rate $4/3$ code for 4 transmit antennas first given in [18].

*Example 4:* Consider the case when $N = 4$, $T = 6$. In this case $K = 16$, $\lambda = 2$, $n = 2$ and $g = 8$. The rate of the resulting code is $4/3$ cspcu and the order of worst-case decoding complexity is $M^{0.5}$. However, using the grouping scheme in [20] we get a decoding complexity of $M^2$. Thus, the new grouping scheme has reduced the decoding complexity considerably. From Propositions 4 and 5, this code can be decoded using both PIC and PIC-SIC decoders to get full diversity. The grouping scheme is: $\mathcal{I}_1 = \{1,2\}$, $\mathcal{I}_2 = \{3,4\}$,...,$\mathcal{I}_8 = \{15,16\}$. For a full-diversity $2 \times 2$ rotation matrix $Q$ we have

$$[z_{2k-1} \ z_{2k}]^T = Q[x_{2k-1} \ x_{2k}]^T \text{ for } k = 1, \ldots, g.$$

TABLE I
COMPARISON OF FULL-DIVERSITY, PIC AND PIC-SIC DECODABLE CODES

| Code | Transmit Antennas $N$ | Delay $T$ | Number of groups $g$ | Number of real symbols per group $\lambda$ | Full diversity with PIC decoding? | Rate in cspcu $R$ | Worst-case Decoding Complexity |
|---|---|---|---|---|---|---|---|
| Toeplitz [13] | $\geq 1$ | $\geq N$ | $2(T-N+1)$ | 1 | Yes | $1 - \frac{N-1}{T}$ | $O(1)$ |
| Code in [18] ($\mathcal{C}_1$) | 2 | 3 | 4 | 2 | Yes | $4/3$ | $M^{0.5}$ |
| Codes in [19] | $\geq 1$ | $\geq N$ | $T-N+1$ | $2N$ | Yes if $T \leq N+1$ | $N\left(1 - \frac{N-1}{T}\right)$ | $M^N$ |
| **Codes in Sec III** | $\geq 1$ | $\geq N$ | $2(T-N+1)$ | $\leq N$ | Yes if $T \leq N+1$ | $\lambda\left(1 - \frac{N-1}{T}\right)$ | $M^{\frac{\lambda-1}{2}}$ |
| Code in [18] ($\mathcal{C}_2$) | 4 | 6 | 8 | 2 | Yes | $4/3$ | $M^{0.5}$ |
| Codes in [20] | $2m, m \geq 1$ | $2l, l \geq \frac{N}{2}$ | $T-N+2$ | $N$ | Yes if $T \leq N+2$ | $\frac{N}{2}\left(1 - \frac{N-2}{T}\right)$ | $M^{\frac{N}{2}}$ |
| **Codes in Sec IV** | $2m, m \geq 1$ | $2l, l \geq \frac{N}{2}$ | $2(T-N+2)$ | $\frac{N}{2}$ | Yes if $T \leq N+2$ | $\frac{N}{2}\left(1 - \frac{N-2}{T}\right)$ | $M^{\frac{N-2}{4}}$ |

The resulting STBC is

$$\begin{bmatrix} z_1 + iz_3 & z_5 + iz_7 & 0 & 0 \\ -z_5 + iz_7 & z_1 - iz_3 & 0 & 0 \\ z_9 + iz_{11} & z_{13} + iz_{15} & z_2 + iz_4 & z_6 + iz_8 \\ -z_{13} + iz_{15} & z_9 - iz_{11} & -z_6 + iz_8 & z_2 - iz_4 \\ 0 & 0 & z_{10} + iz_{12} & z_{14} + iz_{16} \\ 0 & 0 & -z_{14} + iz_{16} & z_{10} - iz_{12} \end{bmatrix}.$$

*Example 5:* Consider the case when $N = 8$ and $T = 12$. In this case, full-diversity is achieved with PIC-SIC decoding. The rate of the code is 2 cspcu and the worst-case decoding complexity is $M^{1.5}$. On the other hand, the grouping scheme in [20] gives a decoding complexity of $M^4$. Delay optimal, full-diversity rate 2 codes in [12] and [25] have decoding complexities of the order of $M^{10}$ and $M^{9.5}$ respectively. The codes reported in [12] and [25] use the optimal, i.e., ML decoder, whereas the code reported in this paper uses only the suboptimal PIC-SIC decoder. Further, the code reported in [12] has the non-vanishing determinant property. Thus, the codes of this section trade performance to get superior decoding comforts.

*Example 6:* Let $N = 6$, $T = 12$. In this case, we get a rate 2 code with worst-case decoding complexity of $M$. On the other hand, the rate 2 fast-ML-decodable code for 6 antennas reported in [12] has a decoding complexity of the order of $M^8$. The code given in [12] is delay optimal and has non-vanishing determinant property whereas the new code gives enormous reduction in decoding complexity without compromising full diversity.

*D. Comparison of full-diversity PIC and PIC-SIC decodable codes*

Table I gives a summary of comparison of full-diversity PIC and PIC-SIC decodable codes available in literature. Here, $M$ is the size of the underlying complex constellation. The class of codes constructed in Section III of this paper includes a family of codes from [19] together with a new grouping scheme, the Toeplitz codes [13] and the two antenna code of [18]. The class of codes in Section IV includes the codes in [20] together with a new grouping scheme and the 4 antenna code from [18]. Consider the subclass of codes in Section III with $\lambda = N/2$. The worst-case decoding complexity of these codes is $M^{\frac{N-2}{4}}$, same as that of codes of Section IV. However, for identical delay $T$, the rate of the codes in Section IV is $\frac{N}{2}\left(1 - \frac{N-2}{T}\right)$, which is slightly more than the rate $\frac{N}{2}\left(1 - \frac{N-1}{T}\right)$ of the codes in Section III. Codes in Section III can give higher rates at the cost of higher decoding complexity by choosing the parameter $\lambda$ properly. However, the codes in Section IV can have rate at the most $N/2$ cspcu only.

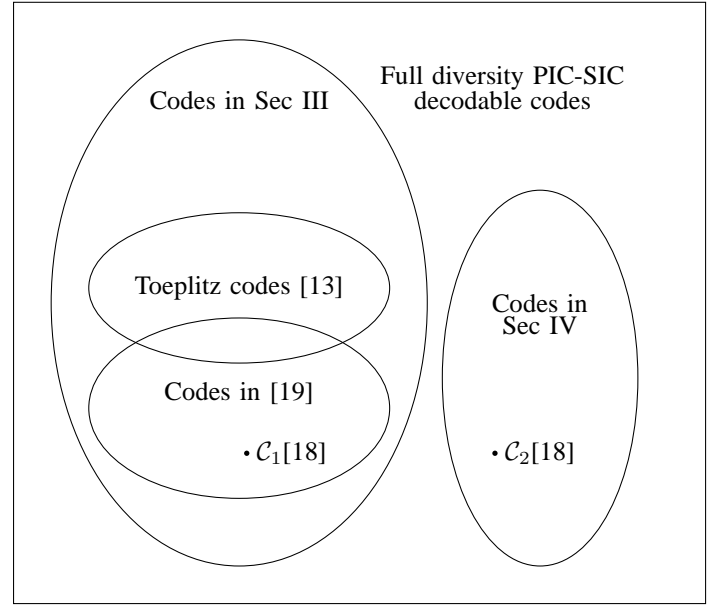

Fig. 1. Venn Diagram of codes listed in Table I

Fig. 1 shows the relationship among the codes listed in Table I. The two antenna code in [18] is denoted by $\mathcal{C}_1$ and the four antenna code of [18] is denoted by $\mathcal{C}_2$. The intersection of Toeplitz codes and codes in [19] corresponds to the subclass of codes in Section III which have $N = 1$. Codes in [20] are exactly the codes in Section IV, but with a different grouping scheme.

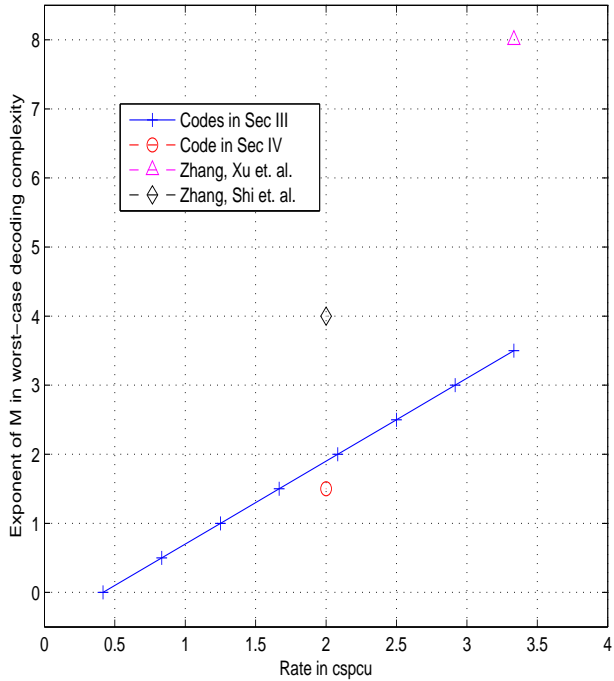

Fig. 2. Comparison of Rate vs. Worst-case-Decoding complexity pairs achievable by various codes for $N = 8$ antennas and delay $T = 12$ channel uses

Fig. 2 shows the comparison of rate vs. worst-case-decoding complexity pairs achievable by the codes in Section III, Section IV, code given by Zhang, Xu et. al. [19] and the code given by Zhang, Shi et. al. [20] for the case of $N = 8$ and $T = 12$. For codes from Section III, each value of $\lambda = 1, \ldots, 8$, gives a different rate-complexity pair. The case $\lambda = 1$ corresponds to a Toeplitz code. The complexity of the codes in Section III is much less than that of the codes from [19] and [20] for identical rates. The code in Section IV and the code in Section III with $\lambda = 4$ have identical worst-case decoding complexity of $M^{1.5}$, however, the code in Section IV has a slightly larger rate. In all other cases, codes from Section III have the best rate-decoding complexity tradeoff.

## V. DISCUSSION

In this paper, we give alternative criteria for STBCs to achieve full diversity with PIC and PIC-SIC decoding. Using the new criteria we constructed a new class of full diversity PIC-SIC decodable codes and we also showed that some of the PIC-SIC decodable STBCs available in the literature can be decoded with lower complexities by choosing the grouping scheme intelligently. The following are some of the directions for future work.

1) Theorems 2 and 4 deal with the full diversity condition only. What is the condition to maximize the coding gain?
2) What is the rate-decoding complexity tradeoff of STBCs with PIC and PIC-SIC decoding?
3) Given an STBC, how does one find the grouping scheme with least decoding complexity and full diversity?

ACKNOWLEDGEMENT

This work was supported partly by the DRDO-IISc program on Advanced Research in Mathematical Engineering through a research grant, and partly by the INAE Chair Professorship grant to B. S. Rajan.